\documentclass[debug,overfull,figures,info,cite]{epl}

\usepackage{amsmath}
\usepackage{graphicx}
\usepackage{amssymb}
\usepackage{bm}

\newcommand{\be}{\begin{equation}}
\newcommand{\ee}{\end{equation}}
\newcommand{\bea}{\begin{eqnarray}}
\newcommand{\eea}{\end{eqnarray}}

\title{Superconductivity Driven by Chain Coupling and Electronic Correlations}
\author{J. M. P. Carmelo\inst{1}\thanks{E-mail: \email{carmelo@fisica.uminho.pt}}
\and F. Guinea\inst{2} \and K. Penc\inst{3} \and P. D.
Sacramento\inst{4}} \shortauthor{J. M. P. Carmelo \etal}
\institute{ \inst{1} GCEP-Center of Physics, U. Minho, Campus
Gualtar, P-4710-057 Braga, Portugal\\
\inst{2} Instituto de Ciencia de Materiales, C.S.I.C.,
Cantoblanco, E-28949 Madrid, Spain\\
\inst{3} Res. Inst. for Solid State Physics
and Optics, H-1525 Budapest, P.O.B. 49, Hungary\\
\inst{4} CFIF - Instituto Superior T\'ecnico, Av. Rovisco Pais,
1049-001 Lisboa, Portugal}

\begin{document}

\maketitle

\begin{abstract}
We present an analysis of a system of weakly coupled Hubbard
chains based on combining an exact study of spectral functions of
the uncoupled chain system with a renormalization group method for
the coupled chains. For low values of the onsite repulsion $U$ and
of the doping $\delta$, the leading instability is towards a
superconducting state. The process includes excited states above a
small correlation pseudogap. Similar features appear in extended
Hubbard models in the vicinity of commensurate fillings. Our
theoretical predictions are consistent with the phase diagram
observed in the (TMTTF)$_2$X and (TMTSF)$_2$X series of organic
compounds.
\end{abstract}

Different classes of materials such as (i) high-$T_c$
superconductors, (ii) quasi-one dimensional (1D) organic
conductors, and (iii) heavy-Fermion compounds show
superconductivity (SC) phases whose physics is far from being well
understood \cite{mat}. All these materials have as common feature
the existence of non-Fermi liquid (NFL) metallic phases. In
addition, in classes (i) and (ii) electronic transport is quite
anisotropic, space dimensionality being expected to play an
important role in the observed exotic spectral properties. On the
other hand, SC phases were not observed in incommensurate quasi-1D
materials \cite{Claessen,branches0}.

The Hubbard model has attracted much interest as the simplest
model for the description of the electronic properties of these
{\it low-dimensional} materials. While the 1D Hubbard model has no
SC phases for on-site repulsion $U>0$ \cite{Carmelo97a}, there has
been evidence that the two-dimensional (2D) Hubbard model has a
$d$-wave pairing instability \cite{dwsc}. In addition, the undoped
1D Hubbard model is the prototype used in the study of the
confinement problem \cite{A87}. Therefore, the study of the
interplay of the crossover from one- to higher-dimensional systems
with the electronic correlations is of great importance for the
understanding of the exotic physics detected in the above
materials \cite{mat}. An interesting toy model for that important
problem is a system of $N_{ch}\rightarrow\infty$ weakly coupled
Hubbard chains, each chain having $N_a\rightarrow\infty$ sites and
the hopping integrals such that $t_{\parallel} > t_{\perp}$. Here
$t_{\parallel}$ and $t_{\perp}$ correspond to electron transfer
along each chain and between first-neighbor chains, respectively.
Importantly, we find below that for electronic doping
concentrations $\delta\neq 0$, where $\delta\equiv (1-n)$ and $n$
is the electronic density, the singlet SC spectral functions of
the one-chain problem show singularities for energy values just
above a correlation pseudogap, $E_{uhb}$, which equals the Mott
Hubbard gap $E_{MH}$ for $\delta\rightarrow 0$ \cite{II}. Note
that half filling for the coupled-chain system could correspond to
a small effective doping $\delta$ for the single chains, due to
charge fluctuations introduced by the inter-chain hopping
\cite{Leo}. Indeed, the key point is that for low values of
$U/t_{\parallel}$ and of doping $\delta$ there is a regime where
$E_{uhb}\approx \vert\delta\vert\,2\pi t_{\parallel}+
[8\sqrt{Ut_{\parallel}}/\pi]\,e^{-2\pi t_{\parallel}/U}$ is small
compared to $t_{\parallel}$ and one can choose $t_{\perp}$ such
that $E_{uhb}<t_{\perp}<t_{\parallel}$. In such a regime the above
finite-energy singularities of the one-chain problem give rise to
relevant instabilities in the system of weakly coupled chains.
Therefore, a complete study of the instabilities of the coupled
chains having as starting point the properties of the one-chain
problem must take into account both the low- and finite-energy
singularities in the one- and two-electron spectral functions of
the 1D problem.

\begin{figure}
\onefigure[scale=0.4]{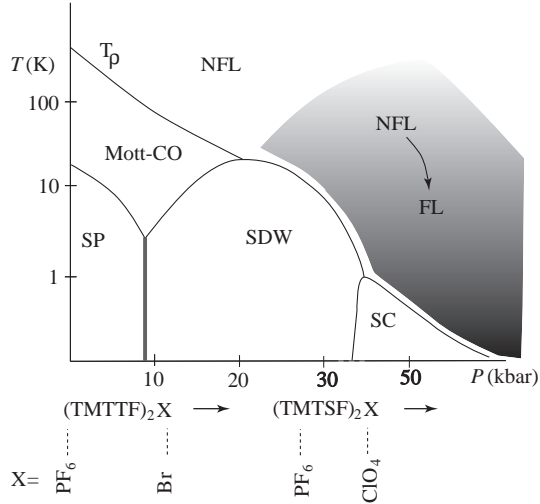} \caption{\label{fig1} T,\,P
Phase diagram of the (TMTTF)$_2$X and (TMTSF)$_2$X series of
compounds.}
\end{figure}

In this Letter we combine the {\it pseudofermion dynamical theory}
(PDT) for these spectral functions
\cite{branches0,theory,branches} with a renormalization group (RG)
analysis \cite{Carmelo97a,chainsRG1,CB} to study the instabilities
of the above system of weakly coupled Hubbard chains. For the
(TMTTF)$_2$X and (TMTSF)$_2$X series of organic compounds, whose
phase diagram is represented in Fig. 1, we identify $E_{uhb}$ with
the experimental {\it correlation gap} \cite{Karlo94} of Fig. 3 of
Ref. \cite{Leo}. Above the energies/temperatures of the
inter-chain hopping and electron-phonon interactions, the 1D
electronic correlations are expected to dominate. Indeed, the 1D
Hubbard model has a NFL metallic phase for $\delta\neq 0$ and a
Mott-Hubbard insulator phase for $\delta =0$, consistently with
the occurrence of NFL and Mott-CO phases in the diagram of Fig. 1.
This model was used successfully in the study of the photoemission
spectrum of the incommensurate organic compound TTF-TCNQ
\cite{Claessen,branches0,branches,Eric}. The description of the
above quarter-filled compounds involves extended Hubbard chains.
However, due to dimerization \cite{Karlo94} the low-$T$ problem
can be mapped onto our toy model for half filling, with $U$ being
an effective interaction parameter whose value depends on the bare
on-site and first-neighbor ($V$) repulsions. Importantly, for
energy and/or $T$ values just above the dimerization gap, this
leads to a correct description also for the small-$\delta$
metallic phase, provided that in our spectral-function expressions
$\delta$ is replaced by an effective doping $\bar{\delta}\equiv
(1/2-n)$ relative to quarter filling. (For instance, for
energy/frequency $(\omega -E_{MH})>0$ small, the optical
conductivity reads ${\rm Re}\,\sigma (\omega)\propto (\omega
-E_{MH})^{1/2}$ for both the $\delta=0$ 1D Hubbard model
\cite{optical} and the ${\bar\delta}=0$ spinless fermion model
with first-neighbor repulsion $V$ \cite{Dias}.) We include in the
perturbative expansion the inter-chain hopping $t_{\perp}$. In
this respect our method differs from alternative approaches where
the expansion parameter used is the electron-electron interaction
$U$ \cite{chainsRG2}. The latter perturbation scheme on $U$ misses
the effects that we discuss here. The influence of the inter-chain
hopping terms has been studied by an expansion in the inverse of
the chain coordination number \cite{chains2}.

The single Hubbard chain in a chemical potential $\mu$ reads
$\hat{H}={\hat{H}}_{SO(4)}-\mu [N_a-N]$, where ${\hat{H}}_{SO(4)}
=\hat{T}+U\sum_{j}[\hat{n}_{j,\uparrow}-1/2][\hat{n}_{j,\downarrow}-1/2]$,
$\hat{T}=-t_{\parallel}\sum_{j,\sigma}[c_{j\sigma}^{\dag}
c_{j+1\sigma} + h. c.]$, $N=\sum_{\sigma}N_{\sigma}$, $N_{\sigma}=
\sum_{j}n_{j,\sigma}$, $c_{j\sigma}^{\dagger}$ and $c_{j\sigma}$
are electron operators of spin projection $\sigma $ at site
$j=1,...,N_a$, and $n_{j,\sigma}=\langle c_{j\sigma }^{\dagger
}c_{j\sigma }\rangle$. $\hat{H}_{SO(4)}$ commutes with the six
generators of the $\eta$-spin and spin algebras
\cite{Yang,theory,II} and hence the $\eta$-spin value $\eta$ and
$\eta_z=-\delta N_a/2$ are good quantum numbers. We consider the
$0<\delta< 1$ metallic and $\delta=0$ insulator phases with zero
magnetization for the initial ground state (GS). The electronic
charge reads $-e$ and $2k_F=(1-\delta)\pi$. We study the spectral
functions $B_{\vartheta }^{(l)}(k,\omega)$ where $l=\pm 1$ and
$B_{\vartheta }^{(+1)}(k,\omega)= \sum_{f}\vert\langle f\vert
{\hat{\cal{O}}}_{\vartheta }^{\dagger} (k) \vert GS\rangle\vert^2
\,\delta (\omega - \omega_{f,0})$ and $B_{\vartheta
}^{(-1)}(k,\omega)= \sum_{f'}\vert\langle f'\vert
{\hat{\cal{O}}}_{\vartheta } (k) \vert GS\rangle\vert^2 \,\delta
(\omega + \omega_{f',0})$ for $\omega >0$ and for $\omega< 0$,
respectively, whose singularities we detect and characterize. The
$f$ and $f'$ summations run over the possible excited states and
$\omega_{f,0}$ and $\omega_{f',0}$ are the energies relative to
the initial GS. We find that the more divergent singular features
occur for one-electron $\vartheta =1p$, transverse spin density
wave (SDW) $\vartheta =sdw$, and on-site and extended singlet SC
$\vartheta =oss$ and $\vartheta =ess$, respectively. The
corresponding operators read ${\hat{\cal{O}}}_{1p} (k)\equiv
c_{k,\sigma}$, ${\hat{\cal{O}}}_{sdw} (k)\equiv
\sum_{k'}c^{\dagger}_{k+k',\downarrow} c_{k',\uparrow}$,
${\hat{\cal{O}}}_{oss} (k)\equiv \sum_{k'}
c_{k-k',\uparrow}c_{k',\downarrow}$, and ${\hat{\cal{O}}}_{ess}
(k)\equiv \sum_{k'}\cos (k')[c_{k-k',\uparrow}c_{k',\downarrow}-
c_{k-k',\downarrow}c_{k',\uparrow}]$. Here
$c_{k,\sigma}^{\dagger}$ and $c_{k,\sigma}$ are spin $\sigma =
\uparrow,\downarrow$ electron operators of momentum $k$. For the
above 1D model, electrons are related to {\it rotated electrons}
by a unitary transformation and {\it rotated-electron double
occupancy} $D_r$ is a good quantum number for all values of the
on-site repulsion $U$ \cite{II,theory}. For $0\leq\delta\leq 1$,
$D_r =0$ for the GS and the first and second upper Hubbard bands
(UHBs) are spanned by $D_r =1$ and $2$ excited states,
respectively \cite{II,theory,branches}, whose lower limits are at
energy $D_r\,E_{uhb}$. The correlation pseudogap $E_{uhb}$ is
called $E_u$ in Ref. \cite{branches} and is such that $E_{uhb} =
4t_{\parallel}\sin (\pi\delta/2)$ for $U/t_{\parallel}\rightarrow
0$, $E_{uhb}=U- 4t_{\parallel}\cos (\pi\delta)$ for
$U>>t_{\parallel}$, $E_{uhb} = U + 4t_{\parallel}$ for
$\delta\rightarrow 1$, and $E_{uhb}=E_{MH}$ for $\delta\rightarrow
0$.

Our study of the above spectral functions $B_{\vartheta
}^{(l)}(k,\omega)$ takes into account all microscopic processes
described by the exact PDT introduced in Ref. \cite{theory}. The
excited states can be described in terms of occupancy
configurations of holons, spinons, and $c$ pseudofermions
\cite{theory,branches}. We denote the $\eta$ spin (and spin) and
$\eta$-spin (and spin) projection of these objects by $s_c$ and
$\sigma_c$ (and $s_s$ and $\sigma_s$), respectively. The holons
(and spinons) have $s_c=1/2$, $\sigma_c=\pm 1/2$, charge $\pm 2e$,
and $s_s=0$ ($s_s=1/2$, $\sigma_s=\pm 1/2$, and no charge degrees
of freedom). The $c$ pseudofermions carry charge $-e$ and have no
spin and $\eta$-spin degrees of freedom. We use the notation $\pm
1/2$ holons (and $\pm 1/2$ spinons) according to the values of
$\sigma_c$ (and $\sigma_s$). The rotated-electron double
occupation $D_r$ equals the number of $-1/2$ holons. In the
vicinity of the singular spectral features the spectral functions
can be expressed as a power-law expansion whose small parameter
for a given value of the momentum is the energy deviation from the
singular feature. For the one- and two-electron spectral functions
only specific types of $\Delta D_r =0$ and/or $\Delta D_r =1$
processes contribute to the leading-order term of such a power-law
expansion. (Here $\Delta D_r =D_r$ is the deviation from the
ground-state value $D_r=0$.) We call these processes {\it dominant
processes}. All remaining processes give rise to higher-order
power-law contributions, controlled by exponents larger than those
studied in this paper. The evaluation of the weight distribution
in the vicinity of the singular features of the above two-electron
spectral functions proceeds as for the one-electron case reported
in Ref. \cite{branches}. The dominant processes involve changes in
the occupancies of the $c$ pseudofermions, two-holon $s_c=0$
composite $c1$ pseudofermions, two-spinon $s_s=0$ composite $s1$
pseudofermions, and localized and non-interacting $-1/2$ Yang
holons, which are not part of composite pseudofermions
\cite{II,theory}. There are no $-1/2$ Yang holons and $c1$
pseudofermions in the $D_r=0$ GS. The $\alpha=c,s1,c1$
pseudofermions carry momentum ${\bar{q}} = q + Q_{\alpha}
(q)/N_a$. Here $q$ is the {\it bare-momentum} and $Q_{\alpha}(q)$
is the phase-shift functional defined in Ref. \cite{theory}, whose
expression involves the bare-momentum two-pseudofermion phase
shifts $\Phi_{\alpha\,\alpha'}(q,q')$ in units of $2\pi$, where
$\alpha,\,\alpha'=c,s1,c1$. (Such phase shifts are studied in Ref.
\cite{branches}.) The pseudofermions are related to the
corresponding {\it pseudoparticles} of Ref. \cite{II} by a unitary
transformation \cite{theory}. The $c1$ energy dispersion
$\epsilon_{c1} (q) = E_{uhb} +\epsilon_{c1}^0 (q)$ such that
$\vert\,q\vert\leq \delta\pi$ plays an important role in our study
and is plotted in Ref. \cite{II}.

Nearly the whole Cooper-pair addition spectral weight is generated
by processes such that $\Delta D_r=0,1,2$. These processes
generate three separated parts $B^{(+1)}_{\vartheta, D_r}
(k,\omega)$ for the SC spectral functions,
$B^{(+1)}_{\vartheta}(k,\omega)
=\sum_{D_r=0,1,2}B^{(+1)}_{\vartheta, D_r} (k,\omega)$. At
$\delta=0$ only $D_r=2$ excited states are allowed and thus
$B^{(+1)}_{\vartheta }(k,\omega)=B^{(+1)}_{\vartheta ,2}
(k,\omega)$ has no singular spectral features. For $\delta
>0$ and $U>0$ the SC singular spectral features result from
transitions to $D_r=1$ excited states. The most divergent
singularities in the SC functions occur for $\vartheta =oss,\,ess$
in $B^{(+1)}_{\vartheta ,1} (k,\omega)$ and result from
finite-energy transitions involving creation of a $c1$
pseudofermion and corresponding to $\omega $ just above the two
$\iota'=\pm 1$ singular branch lines
$E_{ss}^{\iota'}(k)=\epsilon_{c1}(q)$ where $q=[\iota'\,2k_F -
k]$. Such a process involves a $\eta$-spin flip which transforms a
$+1/2$ holon onto a $-1/2$ holon and leads to the deviations
$\Delta \eta =-1$ and $\Delta \eta_z=+1$. By use of the method of
Ref. \cite{branches}, we find $B^{(+1)}_{\vartheta } (k,\omega) =
C^{(+1)}_{\vartheta ,\,\iota'} (k)(\omega
-E_{ss}^{\iota'}(k))^{\zeta_{ss}^{\iota'}(k)}$ for low positive
values of $(\omega-E_{ss}^{\iota'}(k))$ where $C^{(+1)}_{\vartheta
,\,\iota'} (k)$ is real and positive. The exponent $\zeta_{ss}
(k)\equiv\zeta_{ss}^{+1} (k)=\zeta_{ss}^{-1} (-k)$ such that
$\zeta_{ss} (k)=-1 + \sum_{\iota =\pm 1}\{\sqrt{K_{\rho}/2} +
\Phi_{c0\,c1}(\iota\,2k_F,\,q)\}^2 + \sum_{\iota =\pm
1}\{\Phi_{s1\,c1}(\iota\,k_F,\,q)\}^2$ refers to $k\in
([1-2\delta]\pi,\,\pi)$ and involves the phase shifts defined in
Ref. \cite{branches} and the parameter $1/2<K_{\rho}<1$
\cite{Schulz}. The multiplicative constant $C^{(+1)}_{\vartheta
,\,\iota'} (k)$ vanishes for $k=\iota'\pi$ and in the limits
$\delta\rightarrow 0$ and $U/t_{\parallel}\rightarrow 0$ for all
values of $k$. While $C^{(+1)}_{\vartheta ,\iota'} (\iota'\pi)=0$
for the $oss$ and $ess$ $c1$ pseudofermion singular branch lines,
creation of a $-1/2$ Yang holon leads in the metallic phase to a
single $\delta$-function peak for $\vartheta =oss$, $B^{(+1)}_{oss
,1} (\iota'\pi,\omega)=\delta N_a\,\delta (\omega -E_{uhb})$. This
involves a rotation in $\eta$-spin space called {\it
$\eta$-pairing mechanism} \cite{Yang} such that $\Delta \eta =0$
and $\Delta \eta_z=+1$. This peak has a significant fraction of
the $oss$ first UHB weight but does not contribute to the
coupled-chain instability. It is absent for the $ess$ function,
such that $B^{(+1)}_{ess ,1} (\iota'\pi,E_{uhb})=0$. Thus, for the
latter function the whole first-UHB weight corresponds to the
spectral feature generated by creation of the $c1$ pseudofermion.
Consistently, although the $oss$ and $ess$ $c1$ pseudofermion
singularities are controlled by the same exponents, $C^{(+1)}_{ess
,\,\iota'}(k)>C^{(+1)}_{oss ,\,\iota'}(k)$. Therefore, in our
analysis of the coupled chains we do not consider the $oss$
singularity.

The remaining more divergent one-chain singularities were already
calculated in Ref. \cite{Carmelo97a} and for small positive values
of $\omega$ read $B^{(-1)}_{1p }(\mp k_F,-\omega) \propto
B^{(+1)}_{1p }(\pm k_F,\omega) \propto (\omega)^{\zeta_{1p}}$ and
$B^{(-1)}_{sdw}(\mp 2k_F,-\omega)\propto B^{(+1)}_{sdw}(\pm
2k_F,\omega) \propto (\omega)^{\zeta_{sdw}}$. The exponents are
given by $\zeta_{sdw}=-[1-K_{\rho}]$ and $\zeta_{1p}
=-(3/2)\{1-(1/3)[1/(2K_{\rho})+K_{\rho}/2]\}$. Here
$-1/2<\zeta_{sdw}<0$ and $-1<\zeta_{1p} <-7/8$, whereas
$\zeta_{ss} (k)$ is a monotonous decreasing function of $k$ such
that $\zeta_{ss} (k)>0$ for $k=[1-2\delta]\pi$ and $\delta
>0$ and $\zeta_{ss} (k)\rightarrow -1$ as $k\rightarrow\pi$. For
$U>0$, $\zeta_{1p}$ and $\zeta_{ss} (k)$ are a decreasing and
increasing function of $\delta$, respectively, and
$\zeta_{1p}=-7/8$ and $\zeta_{ss} (k)= -1$ for $\delta\rightarrow
0$. The regime of interest for our problem corresponds to small
values of $\delta$ and $U/t_{\parallel}$ and of $[\pi -\vert
k\vert ]>0$ for $\zeta_{ss} (k)$, when $E_{ss}^{\iota'}(k) \approx
2t_{\parallel}[\pi -\vert k\vert] + E_{uhb}$. The correlation
pseudogap $E_{uhb}$ is small for $\delta\in (0,\,0.02)$ and
$U/t_{\parallel}\in (0,\,2)$, when $E_{uhb}\approx \delta\,2\pi
t_{\parallel}+ [8\sqrt{Ut_{\parallel}}/\pi]\,e^{-2\pi
t_{\parallel}/U}<<t_{\parallel}$ and the condition $E_{uhb}
<t_{\perp}<t_{\parallel}$ is fulfilled. The exponents
$\zeta_{ss}(k)$ for $k=(1-\delta/4)\pi$ and $\zeta_{1p}$ are
plotted in Fig. 2 as a function of $\delta $ for
$U/t_{\parallel}=1.0$ and $1.5$. (The exponent $\zeta_{sdw}>-1/2$
for $U>0$ does not appear in the figure.) For low values $\delta$
below some value $\delta_0 = \delta_0 (U/t_{\parallel})$ the
exponent $\zeta_{ss}$ is the smallest and for $\delta>\delta_0$,
$\zeta_{1p}$ is the smallest exponent. Let $k=(1-\delta/a)\pi$
with $a>1$ in the argument of $\zeta_{ss}(k)$. For the same value
of $U/t_{\parallel}$ (and $a$), $\delta_0$ is an increasing
function of $a$ (and $U/t_{\parallel}$). Thus, the same
qualitative results are reached for smaller values of
$U/t_{\parallel}$ and $\delta_0$, when the inequality
$E_{uhb}<<t_{\perp}<<t_{\parallel}$ is fulfilled.

\begin{figure}
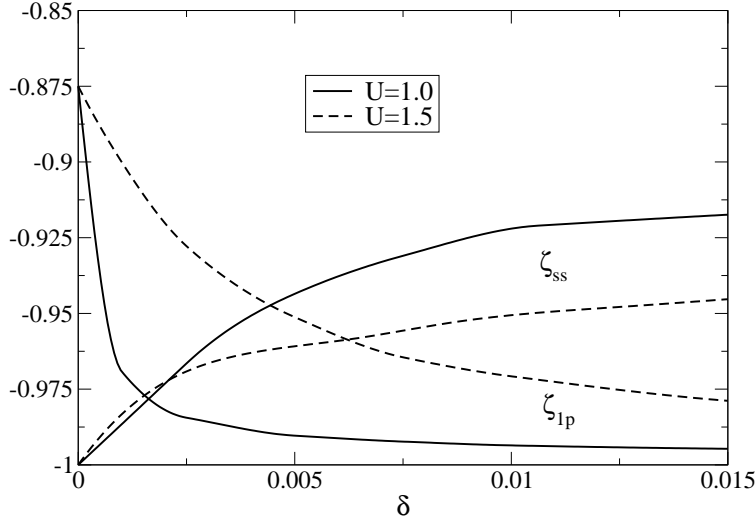

\onefigure[scale=0.4]{CHFig2.eps} \caption{\label{fig2} The
exponents $\zeta_{1p}$ and $\zeta_{ss}([1-\delta/4]\pi)$ as a
function of $\delta$ for $U=t_{\parallel}$ (full line) and
$U=1.5\, t_{\parallel}$ (dashed line).}
\end{figure}

The renormalization of inter-chain couplings can be directly
expressed in terms of the exponents which characterize the
intra-chain spectral functions. For the 1D system the most
divergent singularities occur in the $1p$ and $ess$ functions and
then we can restrict our attention to terms of the type,

\begin{eqnarray}
{\hat{H}}_{Jc} & = & \sum J\, f ( k' ) f ( k'' )
c^{\dag}_{i,k-k',\uparrow} c^{\dag}_{i,k',\downarrow}
c_{i',k-k'',\uparrow} c_{i',k'',\downarrow} \, ; \nonumber \\
{\hat{H}}_{ph} & = & \sum t_{\perp} c^{\dag}_{i,k,\sigma}
c_{i',k,\sigma}  \, . \label{couplings}
\end{eqnarray}
Here ${\hat{H}}_{Jc}$ and ${\hat{H}}_{ph}$ refer to Josephson
coupling and particle hopping, respectively, $f ( k' )$ defines
the extended $s$ pairing, and $i$ and $i'$ denote chain indices.
These couplings are relevant and grow at a rate determined by the
RG equations: ${\partial \tilde{t}_{\perp}}/{\partial l} = -
\zeta_{1p}\, \tilde{t}_{\perp}$ and ${\partial
\tilde{J}}/{\partial l} =- \zeta_{ss}\, \tilde{J} + {\cal C}\,
\tilde{t}_{\perp}^2$. Here $l = -\log ( \Lambda / \Lambda_0)$,
$\Lambda_0$ is the initial value of the cutoff, $\tilde{t}_\perp$
and $\tilde{J}$ are the couplings divided by $\Lambda$, and ${\cal
C}$ is a constant of order unity. Initially, the rate of growth of
the Josephson coupling is proportional to $\tilde{t}_{\perp}^2$,
and when this coupling reaches a value, in dimensionless units, of
order $ \tilde{t}_{\perp}^2$, its growth will be determined by the
scaling dimensions. The analysis is valid until one of the
couplings, $\tilde{t}_\perp$ or $\tilde{J}$, becomes of order
unity, or when $\Lambda \sim E_{uhb}$. Integrating the RG flow, we
obtain $\tilde{t}_{\perp} ( l ) = \tilde{t}_{\perp} ( 0 )\, e^{
|\zeta_{1p}| l}$ and $\tilde{J} ( l ) = \tilde{t}_{\perp}^2 ( 0
)\, [{ e^{ 2 |\zeta_{1p}| l} - e^{|\zeta_{ss} | l}}]/[{2
|\zeta_{1p}| - |\zeta_{ss} |}] + \tilde{J}_0 e^{| \zeta_{ss} |
l}$, where $\tilde{J}_0$ is the initial Josephson coupling. Note
that as for spin singlet SC, all four-operator Hamiltonian terms
have couplings whose leading-order term is $\tilde{t}_{\perp}^2 (
0 )\, [e^{ 2 |\zeta_{1p}| l}]/[{2 |\zeta_{1p}| -
|\zeta_{\vartheta} |}]$. However, since $|\zeta_{\vartheta} |$ is
largest for $\vartheta =ss$, the multiplicative constant
$\tilde{t}_{\perp}^2 ( 0 )/[{2 |\zeta_{1p}| - |\zeta_{\vartheta}
|}]$ is also largest for $\vartheta =ss$. This is consistent with
our choice of Hamiltonian terms, given in Eq. (\ref{couplings}).
As $2 |\zeta_{1p}| > |\zeta_{ss} |$, the RG analysis breaks down
at a scale $\log ( \Lambda_{eff} / \Lambda_0 ) \approx -
|\zeta_{1p} |^{-1} \log [ \tilde{t}_{\perp} ( 0 ) ]$. Provided
that $\Lambda_{eff}$ is greater than $E_{uhb}$, the analysis
presented here gives a reasonable description of the properties of
the system at energies higher than $\Lambda_{eff}$. Otherwise, the
scaling equations cease to be valid at $\Lambda \sim E_{uhb}$. Let
us assume that $\Lambda_{eff} \gg E_{uhb}$. At this scale both
$\tilde{t}_{\perp}$ and $\tilde{J}$ become of order unity. Thus,
the system presents two regimes: (i) at high energies or
temperatures, greater than $\Lambda_{eff}$, the system is best
described in terms of decoupled chains: the effective inter-chain
coupling, in absolute units, decreases upon lowering the
temperature or energy, as $t_{\perp} ( T ) \sim t_{\perp}
(\Lambda_0  )\, ( T / \Lambda_0 )^{1 - | \zeta_{1p} |}$,
inter-chain pair fluctuations increase, and the coherent
contribution to the conductivity decreases; (ii) At sufficiently
low energies or temperatures, as compared to $\Lambda_{eff}$, the
leading terms in the effective Hamiltonian are a one-electron and
a pair-hopping term of similar magnitude, $\sim \Lambda_{eff}$.

Note that the fact that the Josephson coupling has the largest
scaling exponent ensures that its value at the point where the RG
breaks down is the largest among all possible interchain
interactions. The ground state is a strongly coupled
superconductor which will show deviations from the BCS theory. The
low-$T$ phase can be viewed as realization of the inter-layer pair
hopping model derived from coupled systems with strong RVB
correlations \cite{A97}. Our analysis helps to delimit the region
of validity of this hypothesis: (I) $t_{\perp} \ll t_{\parallel}$;
(II) $\delta $ or $\bar{\delta}$ close to zero ; (III) $E_{uhb}$
should be smaller than $t_{\perp}$. Note that the states which
contribute to the pair-hopping instability change the $\eta$
value, {\it i.e.} the initial GS and the excited states are not
related by a mere rotation in $\eta$-spin space. Thus, the present
pairing mechanism cannot be considered directly related to the
processes analyzed in the SO(5) theory of the formation of
$d$-wave SC in the cuprates \cite{Z97}. The study of Ref.
\cite{CB} also leads to singlet SC, but does not use the
singularities of the uncoupled system. Our results are universal
in the sense that from the RG scheme used here we can infer that
the low-$T$ phase of the coupled-chain system will show long-range
SC order. However, the precise nature of this phase, and the
symmetry of the order parameter is dependent on the arrangement of
the chains within the material. Our results confirm that in 1D
NFLs weak inter-chain hopping can induce SC and are consistent
with the phase diagram of the series of quasi-1D compounds of Fig.
1. At low $T$ these materials show a spin Peierls (SP) or SDW
phase. Under pressure the (TMTSF)$_2$X compounds are driven to a
SC phase, which is removed again if one further increases the
pressure. Consistently, for low values of $\delta$ we find that
$E_{uhb}>t_{\perp}$ away from the small $U/t_{\parallel}$ region
and according to the RG scheme of Ref. \cite{Carmelo97a} the
leading instability is towards a SDW state. Increasing pressure
means here increasing $t_{\parallel}$ and $t_{\perp}$ without
changing both $U$ and the condition $t_{\perp}/t_{\parallel}<<1$.
This drives the system into a small-$U/t_{\parallel}$ region,
where $E_{uhb}$ becomes smaller and $t_{\perp}$ is larger, and
thus $E_{uhb}<t_{\perp}$, and the leading instability is towards a
SC state. Since in the above $B^{(+1)}_{ess}(k,\omega)$ expression
$C^{(+1)}_{ess}(k)$ decreases with decreasing $U/t_{\parallel}$
and vanishes as $U/t_{\parallel}\rightarrow 0$, further increase
of pressure removes this state. In the absence of pair hopping,
the renormalization of the inter-chain conductance ceases at
$\Lambda_{eff}$, leading to an effective Fermi-liquid (FL)
description at low $T$. The ability of our toy model to describe
the diagram of Fig. 1 for $X=ClO_4$ can be further checked by
deriving the values for its effective parameters from the NFL
phase for $T\approx 10\,K$ and verifying whether they are
consistent with the occurrence of the SC phase. The use of the PDT
for the uncoupled system leads to a $\omega$-dependent
conductivity for the NFL phase including a small $\omega =0$ Drude
peak and a finite-energy absorption for
$E_{opt}<\omega<E_{opt}+8t_{\parallel}$ where
$E_{opt}=E_{uhb}+\epsilon^0_{c1}(0)$. This absorption is generated
by creation of a $-1/2$ holon and two $c$ pseudofermion holes at
$q$ and $-q$ for $\vert q\vert\in (0,\,2k_F\approx\pi)$. As found
in Ref. \cite{optical} by use of a preliminary version of the PDT,
quantitative agreement with the $X=ClO_4$ and $T=10\,K$ curve of
Fig. 1(C) of Ref. \cite{Leo} is obtained provided that
$t_{\parallel}\approx 0.125\,eV$, $U/t_{\parallel}\approx 1.5$,
and $\delta\approx 0.005$. Importantly, according to Fig. 2, these
values are consistent with the occurrence of the SC phase for
smaller $T$. Tight-binding model (TBM) calculations lead to
$t_{\parallel}\approx 0.250\,eV$ \cite{Leo} because the width of
the $c$ energy-band filled sea for $\vert q\vert <2k_F\approx\pi$
is about twice that of the naive TBM band \cite{II}. Our mechanism
also explains the absence of SC phases in quasi-1D materials whose
$\delta$ value is not close to commensurability. For TTF-TCNQ,
$\delta=0.41$ \cite{Claessen,branches0,branches}, the correlation
pseudogap $E_{uhb}$ is finite for all values of $U$, and thus the
condition $E_{uhb}<t_{\perp}$ cannot be fulfilled. Although our
analysis refers to $t_{\parallel}
> t_{\perp}$, our results suggest that upon increasing
$t_{\perp}/t_{\parallel}$ the SC instability will be extended to
larger values of both $\delta$ and $U$. While for the quasi-1D
system the instability corresponds to momentum values in the
vicinity of $\pi$, we expect that for the high-$T_c$
superconductors square lattice such an instability will appear in
the vicinity of $[\pi,\,\pi]$ as
$t_{\perp}/t_{\parallel}\rightarrow 1$.

We thank C. Bourbonnais (C.B.), A. H. Castro Neto, R. Claessen,
and L. Degiorgi for illuminating discussions and C.B. for
providing Fig. 1. K.P. thanks the financial support of the OTKA
grant T037451.

\end{document}